\documentclass[11pt]{article}
\usepackage{moriond,epsfig}

\def\Journal#1#2#3#4{{#1} {\bf #2}, #3 (#4)}

\def\PLB{{\em Phys. Lett.}  B}

\def\PRD{{\em Phys. Rev.} D}

\def\ra{\rightarrow}

\def\be{\begin{equation}}
\def\ee{\end{equation}}
\def\bea{\begin{eqnarray}}
\def\eea{\end{eqnarray}}

\begin{document}
\vspace*{4cm}
\title{W/Z + JETS AND W/Z + HEAVY FLAVOR PRODUCTION AT THE LHC}

\author{A.A. PARAMONOV}

\address{for the ATLAS and CMS collaborations\\ 
High Energy Physics Division, Argonne National Laboratory,Argonne, Illinois 60439, USA}

\maketitle\abstracts{The ATLAS and CMS experiments at the LHC conduct
an extensive program to study production of events with a $W^\pm$ or
$Z^0$ boson and particle jets. Dedicated studies focus on final states
with the jets containing decays of heavy-flavor hadrons ($b$-tagged
jets). The results are obtained using data from proton-proton
collisions at $\sqrt{s}=7$ TeV from the LHC at CERN. The set of
measurements constitute a stringent test of the perturbative QCD
calculations.}

\section{Introduction}

\par Production of jets in association with a massive vector boson
($W^\pm$ or $Z^0$) is a well-understood process that provides tests of
calculations based on quantum chromodynamics (QCD). These events are
also substantial backgrounds to standard model (SM) measurements and
searches for new physics. The studies of the associated production
constitute a foundation for development of perturbative QCD (pQCD)
calculations and Monte Carlo (MC) simulations. The ATLAS~\cite{ATLAS}
and CMS~\cite{CMS} experiments at the LHC have reported their results
using data from proton-proton collisions at $\sqrt{s}=7$ TeV
collisions in
Refs.~\cite{Chatrchyan:2011ne,Aad:2011xn,Aad:2011qv,Aad:2012en}.
Previously, the associated production of a massive vector boson and
jets was studied at the Tevatron using $p\bar{p}$ collisions at
$\sqrt{s}=1.96$ TeV. The measurements at the LHC offer wider reach in
momenta of the jets than the previous studies.

\par Production of jets containing heavy-flavor hadrons in association
with a massive boson is of special interest. The results of these
studies are presented in
Refs.~\cite{CMS-PAS-SMP-12-003,Aad:2011kp,Aad:2011jn,CMS-PAS-EWK-11-015,CMS-PAS-EWK-11-013}.
Identification of jets with decays of heavy flavor hadrons,
$b$-tagging, was performed via reconstruction of a secondary vertex
within a jets. In Ref.~\cite{CMS-PAS-EWK-11-015} jets were not used
but $B$-mesons were identified via secondary vertices from $B\ra D+X$
decays. The associated production of heavy-flavor hadrons is less
understood than of that of light particle jets. Therefore, the
experimental input is of key importance for development of the MC
simulations and pQCD calculations. Also, these measurements can
provide constraints on the parton density functions (PDF's).

\par The measurements with a $W^\pm$ boson and a $Z^0$ boson are
complementary. Both final states are sensitive to similar physics
processes but they are different from the experimental point of
view. The experimental signatures of the two bosons are
different. Identification of a $W^\pm$ boson requires a
well-identified lepton (an electron or a muon) and large imbalance of
the vector sum of transverse momenta of all reconstructed objects in
event (missing-$p_{\rm T}$). Identification of a $Z^0$ requires two
oppositely-charged leptons of the same flavor (two electrons or two
muons).

\par All the experimental results have been corrected for all known
instrumental effects and are often quoted is a specific range of jet
and lepton kinematics, similar to the detector acceptance. That is
done to avoid prediction-dependent extrapolation and to facilitate
comparisons with theoretical predictions. Theoretical calculations at
next-to-leading order (NLO) in pQCD are presented for final states
with a vector boson and up to four jets.

\section{Backgrounds and Systematic Uncertainties}

\par Reconstruction of the di-lepton invariant mass allows significant
reduction of backgrounds to events with a $Z^0$ boson. The majority of
observed events are from the associated production of a $Z^0$ and
jets. The irreducible backgrounds are the top quark pair production
($t\bar{t}$), dibosons, and $Wt$. These are estimated using MC
simulations normalized with the measured luminosity and predicted
cross sections. Background with one or two non-prompt (``fake'')
leptons are from events with a $W^\pm$ bosons and associated jets and
multi-jet events, correspondingly. Rates of events with ``fake''
leptons are obtained using control regions in data. The requirement
for a jet with decay of a heavy-flavor hadron enhances the fraction of
events from the $t\bar{t}$ production.

\par Events with a $W^\pm$ boson and jets are produced at a higher
rate than with a $Z^0$ boson. The major background with a non-prompt
lepton is from the multi-jet production. The background is evaluated
using orthogonal control regions in data. The contribution from
multi-jet events is different for the electron and muon decay modes of
$W^\pm$ bosons. Therefore, comparison of the measured cross section
from the two decay modes can provide information of biases related to
the evaluation of the backgrounds. The backgrounds with a prompt
lepton are from $t\bar{t}$ production, dibosons, and events with a
$Z^0$ boson and jet. The top pair production becomes the dominant
background in final states with four or more jets (the jets are
counted when $p_{\rm T}>20$, 25, or 30 GeV). The top pair production
is also substantial for events with a $b$-tagged jet. The top pair
production is the dominant background that limits our ability to
measure cross section for events with a $W^\pm$ and two $b$-jets. The
top background is less prominent for measurements involving a $Z^0$
boson in the final state.

\par The major systematic uncertainties are from the jet energy scale
(JES) calibration and efficiency of $b$-tagging. The uncertainty on
the JES grows rapidly when the absolute value of jet rapidity is above
two.

\section{Results}

\par
\begin{figure}[htb]
\centering
\includegraphics[width=0.43\linewidth]{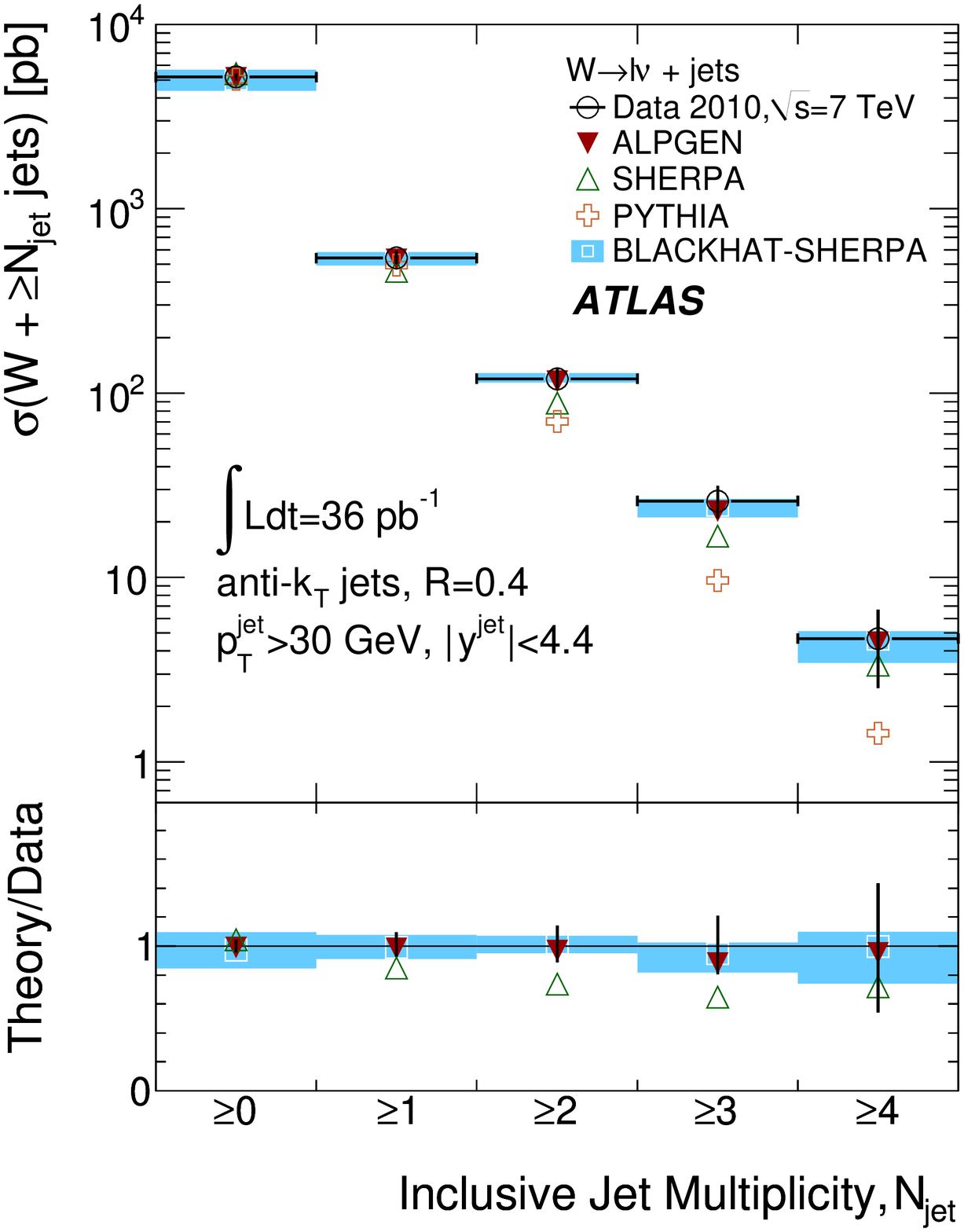}
\includegraphics[width=0.43\linewidth]{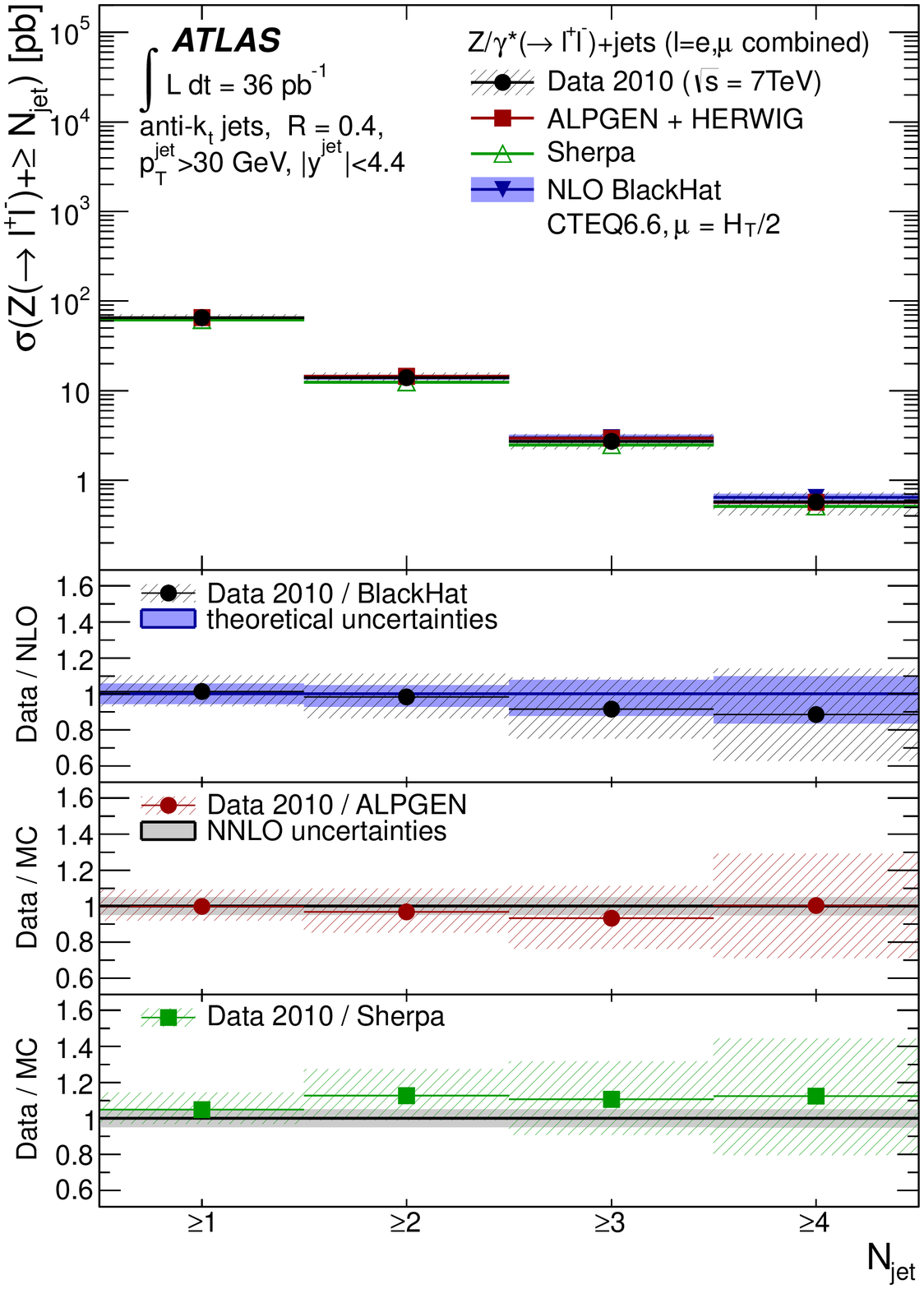}
\caption{Measured cross sections as a function of jet multiplicity for
events with a $W^\pm$ boson~$^6$ (left) and with a $Z^0$ boson~$^5$
(right). The solid bands correspond to the systematic uncertainties on
the predicted cross sections. \label{fig:V_jets}}
\end{figure}

\begin{figure}[htb]
\centering
\includegraphics[width=0.43\linewidth]{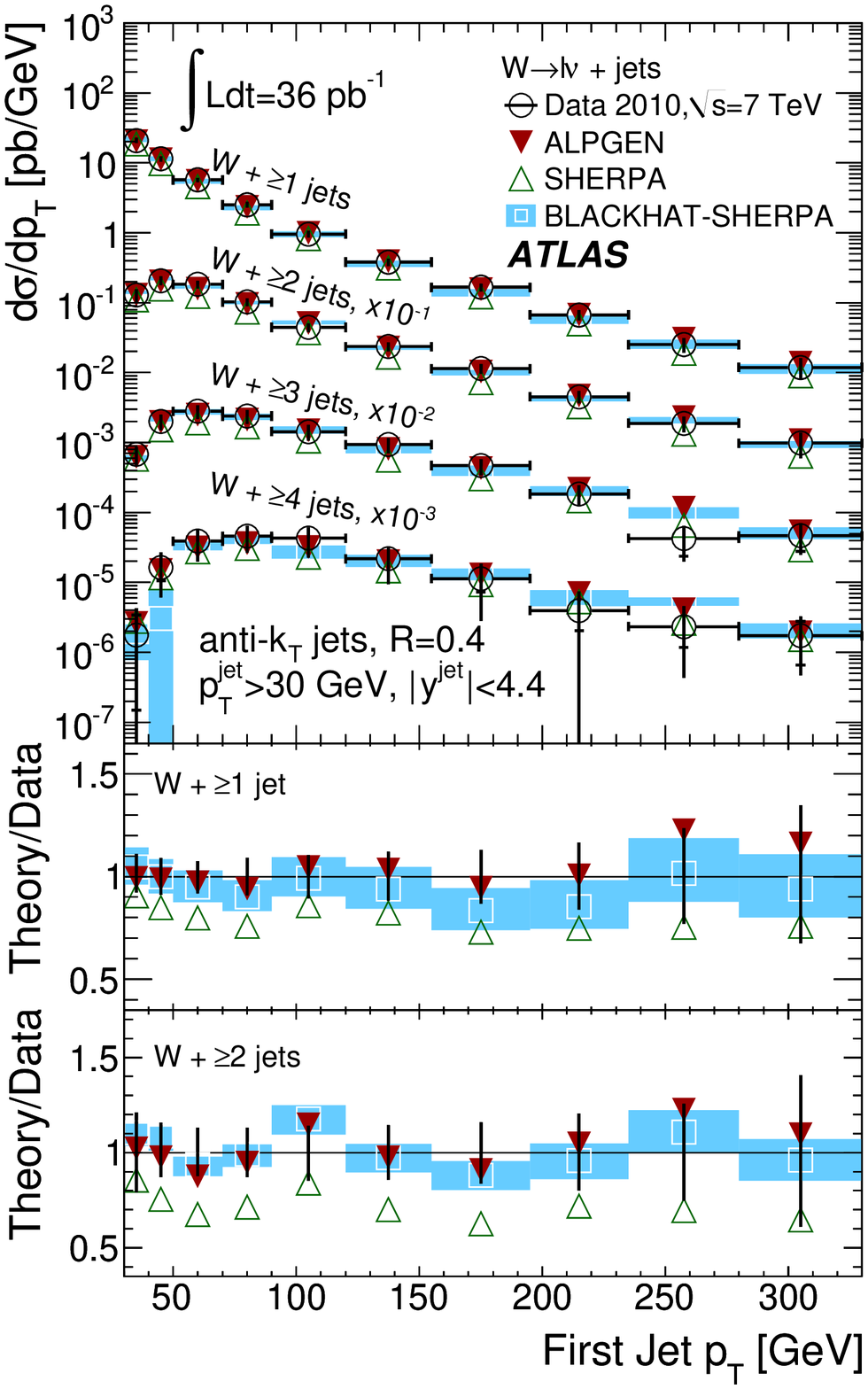}
\includegraphics[width=0.43\linewidth]{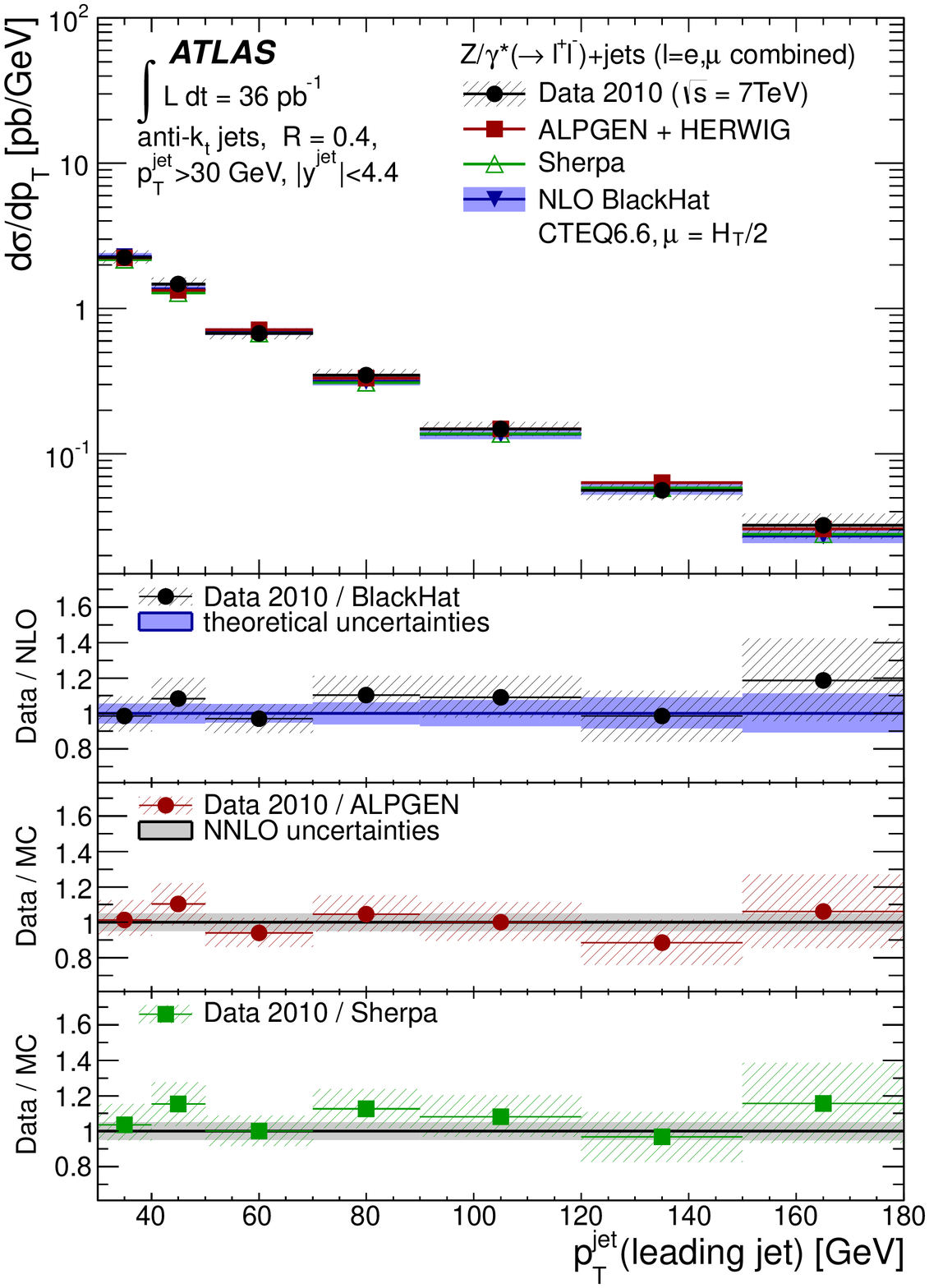}
\caption{Measured cross sections as a function of $p_{\rm T}$ of the
leading jet for events with a $W^\pm$ boson~$^6$ (left) and with a
$Z^0$ boson~$^5$ (right). The solid bands correspond to the systematic
uncertainties on the predicted cross sections. \label{fig:V_jets_pt}}
\end{figure}

 The high cross section of the associated production of a massive
boson and jets allows detailed studies of the kinematic distributions
using differential and inclusive cross sections. Such studies have
been performed by the CMS~\cite{Chatrchyan:2011ne} and
ATLAS~\cite{Aad:2011xn,Aad:2011qv,Aad:2012en} collaborations.
Figs.~\ref{fig:V_jets} and~\ref{fig:V_jets_pt} illustrate the cross
sections measured as a function of inclusive jet multiplicity and
transverse momentum of the leading jet. The studies have been
conducted for a variety of kinematic observables such as invariant
mass of multiple jets, angular and rapidity separation between jets,
and so on. The measured ratios of cross sections allow cancellation of
major systematic uncertainties.

\par The measured cross sections are compared to the NLO calculations
from {\sc BlackHat-Sherpa} and MC simulations from {\sc Pythia}, {\sc
Sherpa} and {\sc Alpgen} matched to {\sc Herwig}. The NLO pQCD
predictions are found in good agreement with data. Leading-order (LO)
matrix element calculations for final states with a vector boson and
up to five partons are matched to parton showering in {\sc Sherpa} and
{\sc Alpgen+Herwig}. These two generators are also in good agreement
with data.

\par Production of a charm hadron in a jet and a $W^\pm$ boson is
reported in Ref.~\cite{CMS-PAS-EWK-11-013}. The study has sensitivity
to the strange quark PDF. Ratios of cross sections were measured to be
$\sigma(W^+\bar{c}+X)/\sigma(W^-c+X)=0.92\pm$0.19(stat.)$\pm$0.04(syst.)
and
$\sigma(Wc+X)/\sigma(W+jet+X)=0.143\pm0.015$(stat.)$\pm$0.024(syst.).
The ratios are measured in the kinematic region $p^{\rm jet}_{\rm
T}>$20 GeV, $|\eta^{\rm jet}|<$2.1 for $W\ra\mu\nu$ decays. The
measured results are in agreement with theoretical predictions at NLO
based on available parton distribution functions.

\par
\begin{figure}[htb]
\centering
\includegraphics[width=0.43\linewidth]{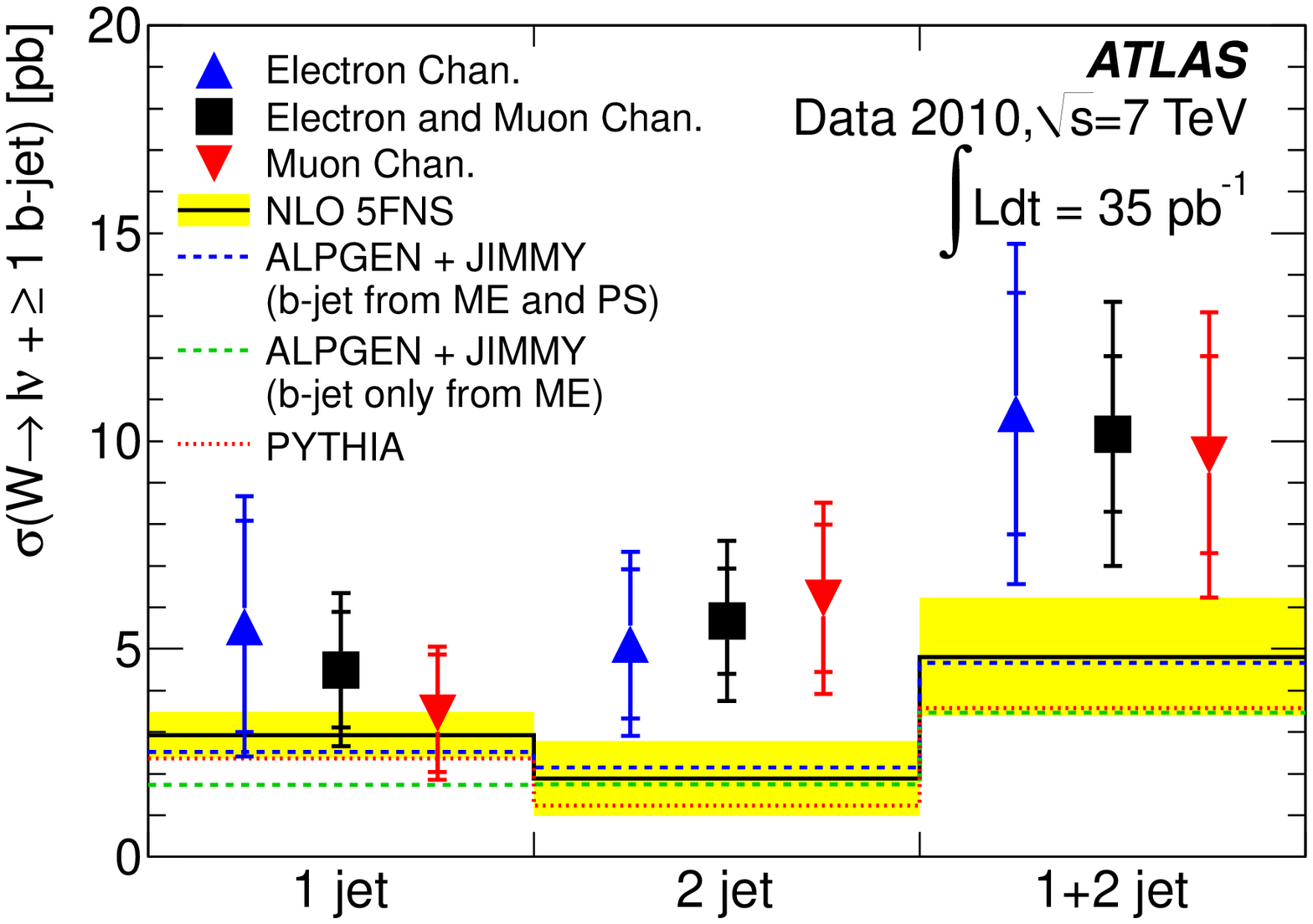}
\includegraphics[width=0.43\linewidth]{FinalDeltaR.epsi}
\caption{Exclusive cross sections for events with a $b$-jet and a
$W^\pm$ (left) from ATLAS~$^8$. Distribution in angular separation,
$\Delta R$, between $B$ meson candidates in events with a $Z^0$
(right) from CMS~$^{10}$. \label{fig:Wbc}}
\end{figure}

 Studies of the associated production of jets with decays of $B$
mesons ($b$-jets) are described in
Refs.~\cite{CMS-PAS-SMP-12-003,Aad:2011kp,Aad:2011jn}. These final
state are backgrounds to the associated Higgs production; $pp\ra HW$
and $pp\ra HZ$, where $h\ra b\bar{b}$. The results for production of a
$b$-jet and a $W^\pm$ boson are presented in Fig.~\ref{fig:Wbc}. The
measured cross section slightly exceeds the predicted value for final
states with a single $b$-jet and another jet.
Ref.~\cite{CMS-PAS-SMP-12-003} presents cross sections for one and two
$b$-jets with $p_{T}^{\rm jet}>25$ GeV and $\eta^{\rm jet}<2.1$. The
measured cross sections are $\sigma(Z^0+\:2\:b{\rm
-jets}+X)=0.37\pm0.02$(stat.)$\pm0.07$(syst.)$\pm0.02$(theory) pb and
$\sigma(Z^0+\:b{\rm
-jet}+X)=3.78\pm0.05$(stat.)$\pm0.31$(syst.)$\pm0.11$(theory) pb. The
cross section for two $b$-jets is in agreement with LO pQCD
predictions.

\par The study of the angular correlations between two $B$ hadrons
produced in association with a $Z^0$ boson is presented in
Ref.~\cite{CMS-PAS-EWK-11-015}. Identification of $B$-hadron
candidates utilizes displaced secondary vertices without involving
jets. That allows to analyze production of $B$ hadrons at small
angular separation. The normalized production cross section as
function of the angular separation is compared with QCD predictions at
tree-level in Fig~\ref{fig:Wbc}. The measurement is performed in the
kinematic region defined for $B$ hadrons with $p_{\rm T}>15$ GeV and
$|\eta|<2$. This study gives further insight into the properties of
heavy quark pair-production in association with a neutral vector
bosons.

%\section*{Acknowledgments}
%\par The author thanks the organizers of the 47th Rencontres de
%Moriond for maintaining the spirit of the conference...

\end{document}